# Simple Expression For Minimum Emittance With Linearly Varied Bending Radius In Dipole Magnets

G. Baranov, E. Levichev and S. Sinyatkin

*Budker Institute of Nuclear Physics, Novosibirsk 630090, Russia*

We study the theoretical minimum emittance for a non-uniform bending magnet with the bending radius linearly ramped from the dipole center to its end. We derive the expression for the minimum emittance as a function of the bending angle and expand it into a power series with respect to a small angle. The first term of the expansion gives the TME minimum emittance while the high-order terms are responsible for its modification. On the contrary of the vague and entangled closed-form solution, the coefficients of the power series are simple and clearly indicate conditions and limitations for emittance reduction below the TME value. With the help of analytical predictions we design a lattice cell with longitudinally varied bends demonstrating the emittance less than that for the TME structure of the same bending angle.

## I. INTRODUCTION

The equilibrium emittance in a relativistic electron storage ring is defined by the balance between radiation damping and quantum excitation and can be expressed as

$$\varepsilon_x = C_q \frac{\gamma^2}{J_x} \frac{I_5}{I_2}, \qquad (1)$$

where $C_q = 3.832 \times 10^{-13}$ m, $\gamma$ is the relativistic factor, $J_x$ is the horizontal damping partition number, and two synchrotron radiation integrals represent damping and excitation, respectively [1]

$$I_2 = \oint_M \frac{ds}{\rho^2(s)}, \qquad I_5 = \oint_M \frac{H(s)ds}{|\rho(s)|^3}. \qquad (2)$$

Here $\rho(s)$ is the curvature radius and the dispersion action is given by

$$H(s) = \gamma_x \eta^2 + 2\alpha_x \eta \eta' + \beta_x \eta'^2, \qquad (3)$$

with the Twiss parameters ($\alpha_x$, $\beta_x$, $\gamma_x$) and the dispersion function and its derivative ($\eta, \eta'$), respectively.

The minimum emittance with uniform bends is achievable in the TME lattice [2 – 6] consisting of a bending magnet with length $L_u$ (the suffix $u$ indicates the uniform magnet to set the corresponding value apart from the varied bend magnet denoted below with the suffix $v$) and bending angle $\theta$, plus number of quadrupoles to adjust the optics. The horizontal beta and dispersion have specified minimum at the middle point of the magnet

$$\beta_{x0u} = L_u / 2\sqrt{15} \qquad \text{and} \qquad \eta_{0u} = \theta \cdot L_u / 24 \qquad (4)$$

which gives the minimum TME emittance of

$$\varepsilon_{xu} = C_q \frac{\gamma^2}{J_x} \frac{\theta^3}{12\sqrt{15}} .$$ (5)

To get over the TME limit, Wrulich in 1992 proposed to use non-uniform magnets with longitudinal variation of the bending field [7]. Horizontal beta and dispersion grow from their minimums at the magnet midpoint toward the magnet ends and cause corresponding increase of the dispersion action $H(s)$ in the quantum excitation integral Eq. (2). According to Wrulich, one can compensate increase of $H(s)$ by enlarging the bending radius (or reduction of the magnet field strength). As a result, additional minimization of $I_5$ is possible for the magnetic field high at the dipole center and low at its edges.

The approach has been intensively elucidated in recent years both analytically and numerically [8 – 13]. Detailed study can be found in trio papers [11 – 13] where a minimum emittance theory was developed for arbitrary dipole bending profile using vector and matrix form. Exact closed-form expression for the minimum emittance with the linearly ramped radius (with the constant field segment) was derived and investigated. The expression (as well as similar ones from other references) is lengthy and cumbersome and the conclusions following from it are obscured by its complexity.

Below we also derive closed-form formulas for the minimum emittance and corresponding initial beta and dispersion in the magnet with linear ramp of the bending radius (hyperbolic field profile). Although these expressions are also entangled, they are functions of the bending angle and for $\theta < 1$ they can be expanded in power series with respect to $\theta$. The first term of the series is the exact TME minimum while the next terms explain possible reduction of the emittance due to the field variation. The terms of the power series are simple, clear and allow making easy predictions of the emittance minimization below the TME limit.

To check validity of the predictions we design the lattice cell with the varied-bend magnet and compare it with the same bending angle TME cell.

## II. TASK DEFINITION

Numerical examination of a TME-like non-uniform magnet reveals the fact that to reach the minimum emittance the curvature radius tends to ramp almost linearly from the low value at the magnet midpoint to the high value at the magnet edge (see, for instance, [8, 12]). Here we study just this case with the bending radius and the field profile demonstrated in Fig.1. No flat top segment is included for simplicity. The magnet center and the end values are $B_c = B_{\max}$ (corresponding to $\rho_c = \rho_{\min}$) and $B_e = B_{\min}$ (corresponding to $\rho_e = \rho_{\max}$), respectively. The length of the varied field magnet is $L_v$. Due to the symmetry we take the magnet center as the reference point and consider only a half of the magnet as it is shown in Fig.1. The bending radius is given by

$$\rho(s) = k \cdot s + \rho_c ,$$ (6)

with the gradient

$$k = \rho' = \frac{2(\rho_e - \rho_c)}{L_{vm}}. \qquad (7)$$

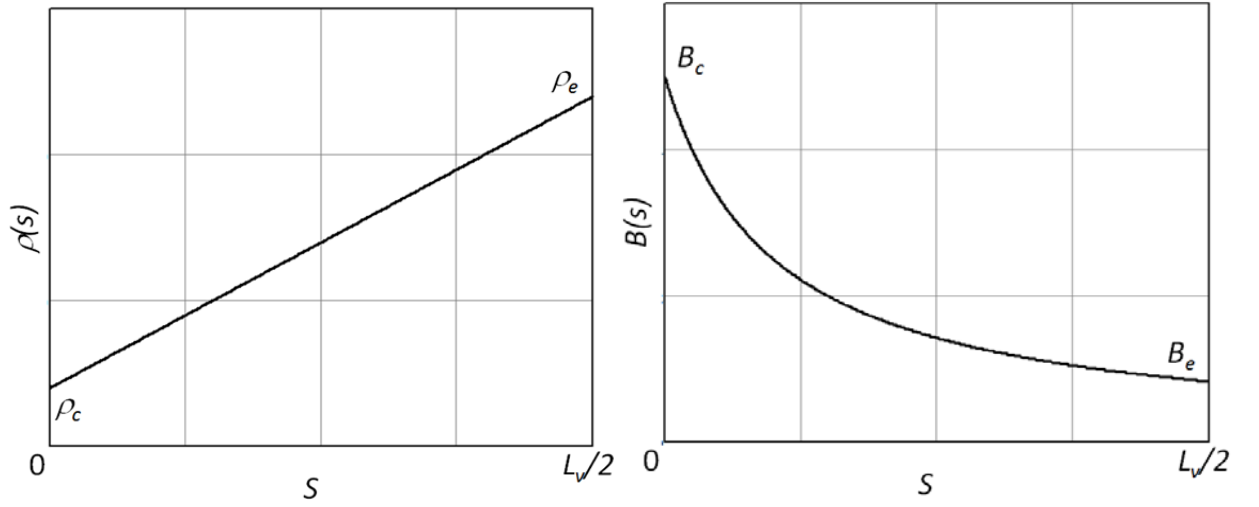

Fig.1 Bending radius and bending field in the magnet half.

Below instead of the orbit length $s$ we use as an independent variable the bending angle

$$\theta(s) = 2k^{-1} \ln\left(1 + \frac{k \cdot s}{\rho_c}\right), \qquad (8)$$

with the total angle over the magnet

$$\theta = 2k^{-1} \ln\left(\frac{\rho_e}{\rho_c}\right). \qquad (9)$$

Let us compare the length of the uniform and non-uniform magnets with the same total bend and maximum field $B_u = B_c$. Inserting Eq. (7) into Eq. (9) we obtain

$$L_v = L_u \frac{x-1}{\ln(x)}, \qquad (10)$$

where $x = \rho_e / \rho_c = B_c / B_e$. As the factor $y = \ln(x) = \ln(B_c / B_e)$ is crucial in the following theory, Fig.2 depicts the elongation of the varied magnet with respect to the constant one.

For the sake of brevity, below we set the horizontal damping partition $J_x = 1$ and omit the factor $C_q \gamma^2$ in emittance (for emittance in nanometers $C_q \gamma^2 = 1468 \cdot E^2(GeV)$). With this notation the minimum TME emittance (5) takes the form

$$\varepsilon_{xu\min} = \frac{\theta^3}{12\sqrt{15}}.$$

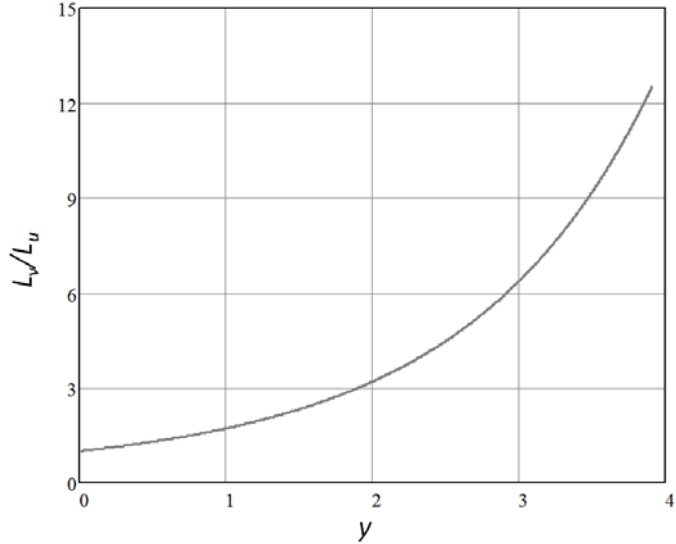

Fig.2 Elongation of a non-uniform magnet relatively to the uniform one vs. $y = \ln(B_c / B_e)$. Both magnets have the same peak field and bending angle.

## III. MINIMUM EMITTANCE

To derive analytically the theoretical minimum emittance for non-uniform magnet with the radius $\rho(s)$ defined by Eq. (6) we follow the conventional procedure. At first, we solve the second-order differential equations for the horizontal betatron motion and for the dispersion function using the initial conditions determined by the symmetry. Then from the betatron transfer matrix we find the horizontal betatron function propagation (which is, actually, almost equal to that for a drift). At the next step we assemble the dispersion action $H(s)$ according to Eq. (3) and calculate two synchrotron radiation integrals Eq. (2). Finally, we minimize the emittance expression Eq. (1) with respect to the initial $\beta_{x0} = \beta_x(0)$ and $\eta_0 = \eta(0)$ (taking into account that due to the reflection symmetry $\beta'_{x0} = \eta'_0 = 0$). In processing we use the bending angle $\theta(s)$ instead of $s$ as an independent variable that looks natural for emittance minimization technique.

The calculations are tedious and were carried out with the help of Mathematica 10.0 computational software program [15]. We expand exact closed-form solution given in Appendix A into a power series for $\theta < 1$ and have the following expression for the emittance

$$\varepsilon_{xv\,min} \approx \frac{\theta^3}{12\sqrt{15}} \cdot \left(1 - \frac{9k\theta}{32} + \frac{2337 k^2 \theta^2}{71680} - \frac{2081 k^2 \theta^3}{1376256} + ...\right). \qquad (11)$$

Here a zero-order term obviously represents the TME constant field case. The next term provides the major contribution to the emittance reduction below the uniform magnet. Replacing $k \cdot \theta = 2y = 2\ln(B_c / B_e)$ we obtain a simple formula

$$\varepsilon_{xv\,min} \approx \varepsilon_{xu\,min} \cdot \left(1 - \frac{9y}{16} + \frac{2337 y^2}{17920} - \frac{2081 y^3}{172032} + ...\right). \qquad (12)$$

In this approximation the emittance decrease depends only on the logarithmic ratio of the maximum and minimum field in the longitudinal gradient bend. Fig.3 shows the emittance reduction factor $r_\varepsilon = \varepsilon_{xv}/\varepsilon_{xu}$ as a function of $k$ for exact formula Eq.(A2) and decomposition Eq. (11) for two bending angles of $\theta = 0.1$ and $\theta = 0.2$.

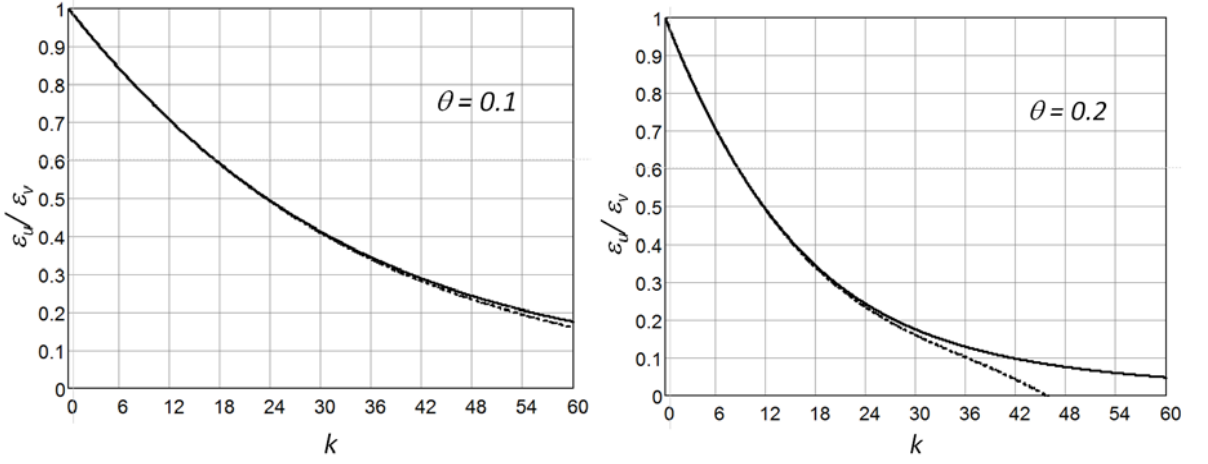

Fig.3 Varied field emittance reduction factor for two bending angles. Solid line corresponds to the exact formula Eq. (A2), dashed line corresponds to the series Eq. (11).

Eq. (12) cannot give us details on the parameters of the uniform TME magnet but the expansions of the initial beta Eq. (A3) and dispersion Eq. (A4)

$$\beta_{xov} \approx \frac{\rho_c \theta}{2\sqrt{15}} \cdot \left(1 + \frac{7k\theta}{32} + \frac{1859 k^2 \theta^2}{215040} - \frac{16501 k^3 \theta^3}{6881280} + \ldots\right), \quad (13)$$

$$\eta_{ov} \approx \frac{\rho_c \theta^2}{24} \cdot \left(1 - \frac{k\theta}{8} - \frac{k^2 \theta^2}{60} + \frac{k^3 \theta^3}{320} + \ldots\right), \quad (14)$$

correspond to the field and the length of the uniform magnet $B_c$ and $L_u = \rho_c \theta$, respectively. Inserting $y$ parameter into Eq. (13) and Eq. (14) yields

$$\beta_{xov} \approx \beta_{x0u}(B_c) \cdot \left(1 + \frac{7y}{16} + \frac{1859 y^2}{107520} - \frac{16501 y^3}{860160} + \ldots\right), \quad (15)$$

$$\eta_{ov} \approx \eta_{0u}(B_c) \cdot \left(1 - \frac{y}{4} - \frac{y^2}{15} + \frac{y^3}{40} + \ldots\right). \quad (16)$$

Fig.4 compares the exact solutions Eqs. (A3, A4) with the power series expansion Eqs. (13, 14) for the total bending angle of $\theta = 0.15$.

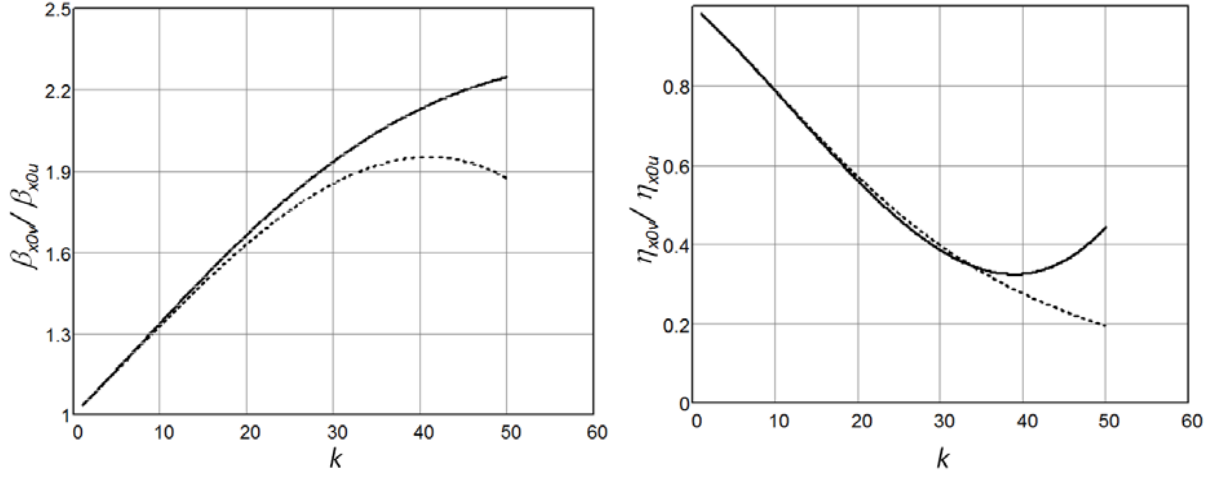

Fig.4 The initial beta (left) and dispersion (right) normalized to the corresponding uniform magnet as a function of the radius gradient $k$. Solid and dashed lines show the exact solutions from Appendix A and the power series expansion respectively.

Although we planned to expand Eqs. (A2, A3, A4) as a power series in small $\theta$, it appeared that the real expansion parameter is $y = k \cdot \theta / 2 > 1$. The detailed study shows that the numerical coefficients in Eqs. (12 – 16) decrease exponentially, but the series convergence depends on the product of $k \cdot \theta$ and for large bending angle and/or large radius gradient convergence can be poor. In this case either the expansions should be extended to the high-order terms or the exact formulas are used.

It is worth noting that the transverse gradient $K = G / B\rho$ in the longitudinally uniform TME magnet also can reduce the minimum emittance below Eq. (5)

$$\varepsilon_{xu\min} \approx \frac{\theta^3}{12\sqrt{15}} \cdot \left(1 - \frac{3(1 + K \cdot \rho^2)\theta^2}{70} + ...\right),$$

but $\theta^2$ for $\theta < 1$ makes this mechanism ineffective compared to $\theta$ in Eq. (11).

## IV. LATTICE CELL DESIGN

To validate theoretical results we attempt to design the minimum emittance lattice cell with linear ramp of bending radius in the dipole. For reference we use the uniform dipole with the constant field $B_c$. The problem is that in a real compact lattice it is difficult to satisfy optimal conditions Eq. (4) so the resulting emittance degrades. So a supplementary issue of the paragraph is how close we can approach the absolute minimum emittance for both constant and varied field TME lattice cell. Solutions provided more compact lattice cell would be an asset as usual.

Also we would like to mention that a systematic design and optimization of the lattice with non-uniform dipole is beyond the scope of this paper, we purpose only to illustrate analytical results of the previous sections.

According to Eq. (12) the only parameter $y = \ln(B_c / B_e)$ defines emittance reduction below the uniform TME magnet. The maximum field $B_c$ is constrained by available technology. For instance, a superconducting dipole with the field ~10 T is referred in [16] and it seems to be close to the practical limit for reasonable vertical aperture. We can increase $y$ by decreasing the edge field $B_e$ but according to Fig.2 it results in lengthening of the varied field magnet with respect to the uniform one. Emittance reduction by an order of magnitude with $y = 4$ (see Fig.A1) elongates the magnet by factor ~14 (see Fig.2). The latter inevitably implies difficulties to the magnet design and causes increase in the storage ring circumference. Fig.5 shows the emittance reduction factor for the linear radius ramp magnet with respect to the uniform one versus the corresponding length increase. Almost linear curve in Fig.5 indicates that for the magnet with linear radius ramp the emittance reduction factor roughly corresponds to that for the orbit elongation.

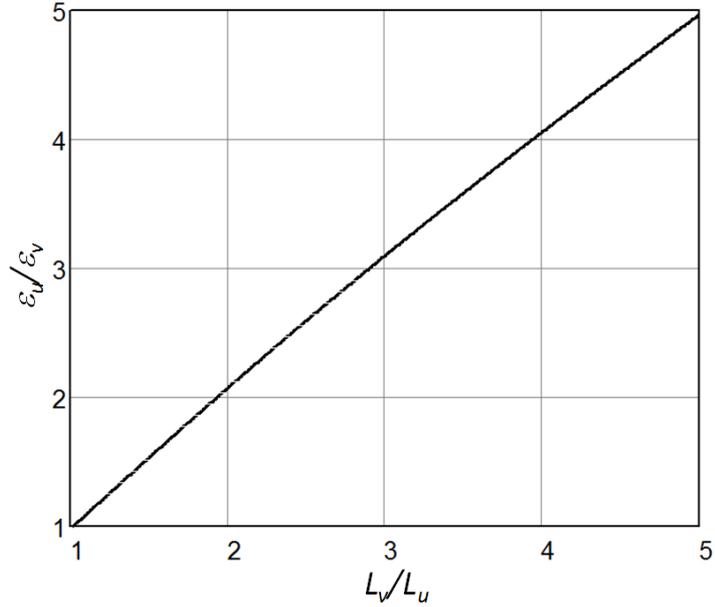

Fig.5 Non-uniform magnet emittance reduction factor vs. the length increase (the maximum field of the uniform and non-uniform magnets is assumed the same).

We start lattice design with a homogenous field TME cell providing $\varepsilon_x = 1\,\text{nm}$ minimum emittance at $E = 3\,\text{GeV}$. We set 10 T maximum field for both magnets. High field gives the length of the varying field dipole convenient for production and saves the machine size. A segmented magnet with a high field compact core and low field wings can also be considered. According to Eq. (5), the optimal bending angle for the minimum TME emittance is $\theta = 0.152$ and the uniform magnet length is $L_u = 0.152$ m. Eqs. (4) give the initial horizontal beta and dispersion at the dipole midpoint.

For the non-uniform dipole we specify the linear radius ramp according to Eq. (6) with $B_c = 10\,\text{T}$ and with the doubled length of the uniform magnet. Fig.5 results the minimum emittance ≈0.5 nm with the initial beta and dispersion defined by Eq. (A3) and Eq. (A4), respectively. The theoretical values are listed in the second column of Table 1.

.

To simulate the minimum emittance we split the dipole with the hyperbolic field profile into a large number of slices and require MAD8 [17] equipped with a simplex optimizer to find the initial beta and dispersion providing the minimum emittance. The optimization results are listed in the third column of Table 1 and correspond well to the theoretical prediction.

Table 1 Uniform and varied field magnets parameters ( $E = 3\,\text{GeV}$, $\theta = 0.152$ )

|  | Uniform | Varied (th) | Varied (op) |
|---|---|---|---|
| $L$, m | 0.152 | 0.304 | 0.304 |
| $\varepsilon_x$, nm | 1.000 | 0.482 | 0.477 |
| $\beta_{x0}$, m | 0.0196 | 0.0305 | 0.0307 |
| $\eta_0$, m | $9.63\cdot10^{-4}$ | $6.19\cdot10^{-4}$ | $6.13\cdot10^{-4}$ |

At the next step we design the simplest TME cell where the central dipole (uniform or non-uniform) is surrounded by two quadrupole doublets. MAD8 with the built-in optimizer was run to find periodic solutions with minimal emittance. A constraint of the maximum quadrupole gradient of ≤40 T/m is imposed on the optimization process. The resulted optical functions are plotted in Fig.6 while some relevant parameters of the cells (marked as "Detuned") are listed in summarizing Table 2.

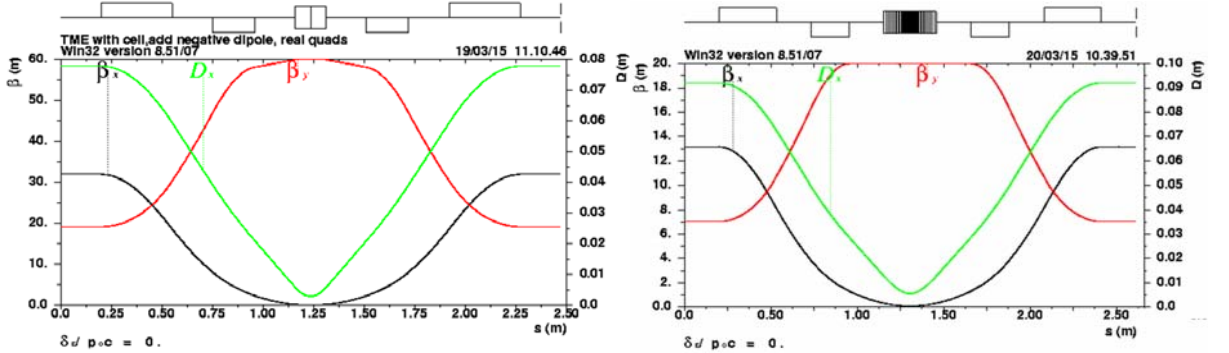

Fig.6 Uniform (left) and non-uniform magnet lattice cell.

The numbers from Table 2 clearly show that with the simplest and compact lattice cell we are very far from the optimum conditions and the minimum emittance. For the uniform dipole we have 2.75 nm instead of 1 nm and for the varied field dipole we have 3.14 nm instead of 0.5 nm. The reason is in poor convergence of the periodical solutions for the horizontal beta and dispersion: there are no control knobs to tune both of them simultaneously. And the situation is even worse for the non-uniform magnet because (see Table 1) the initial beta increases with respect to the uniform magnet while the initial dispersion, on the contrary, decreases.

To improve the periodic optics convergence we, according to the recommendation in [14], introduce a low field negative curvature magnet (anti-bend) into the lattice cell. Varied realizations of the anti-bend are possible. We transfer two quadrupole doublets between the dipoles in Fig.6 to the quadrupole triplet in Fig.7 and insert the anti-bend into the middle defocusing quadrupole.

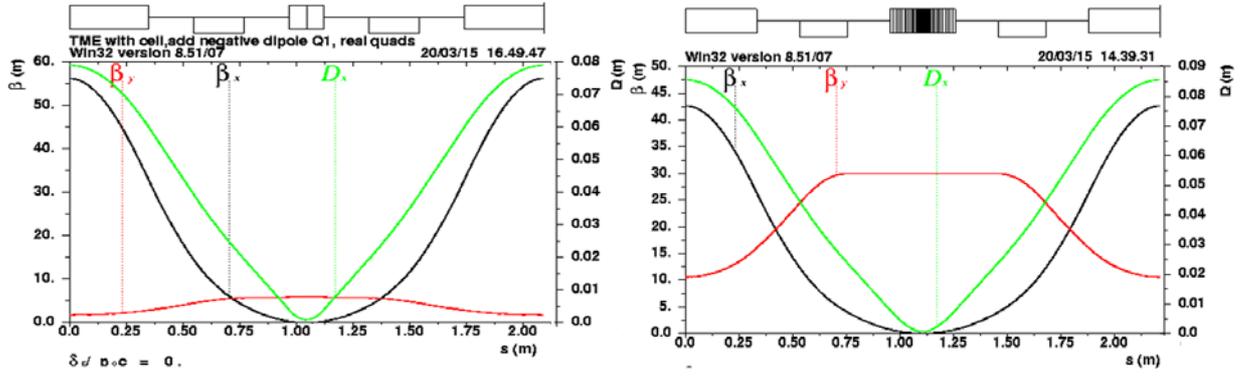

Fig.7 TME cells with the negative field in the central triplet quadrupole.

Rather low field (–61 mT with the full angle of –4.2 mrad in the uniform magnet and –122 mT with the full angle of –8.1 mrad in the non-uniform magnet) is enough to match the cells properly and obtain the theoretical minimum emittance from Table 1. The results are denoted in Table 2 as "Tuned". To produce the required anti-bend, the defocusing quadrupole should be displaced horizontally by 1.5÷3 mm.

Table 2 Results of the low emittance cells design.

| Parameter | Uniform magnet | | Varied radius magnet | |
|---|---|---|---|---|
| | Detuned | Tuned | Detuned | Tuned |
| Length $L$, m | 2.47 | 2.09 | 2.61 | 2.21 |
| Emittance $\varepsilon_x$, nm | 2.75 | 1.07 | 3.14 | 0.53 |
| Initial beta $\beta_{x0}$, $10^{-2}$·m | 3.29 | 1.96 | 10.1 | 3.07 |
| Initial disp. $\eta_0$, $10^{-4}$·m | 29.7 | 9.63 | 55.8 | 6.13 |
| Energy spread $\delta$, $10^{-3}$ | 2.57 | 2.61 | 2.11 | 2.14 |
| Mom.compaction $\alpha$, $10^{-4}$ | 2.41 | –0.013 | 4.60 | –1.03 |
| Energy loss, $10^{-1}$·MeV | 1.73 | 1.83 | 1.09 | 1.10 |
| $I_1$, $10^{-4}$·m | 5.96 | –0.027 | 12.0 | –2.28 |
| $I_2$, $10^{-2}$·m$^{-1}$ | 15.2 | 16.1 | 9.52 | 9.61 |
| $I_3$, $10^{-2}$·m$^{-2}$ | 15.2 | 16.5 | 6.41 | 6.49 |
| $I_4$, $10^{-4}$·m$^{-1}$ | 5.96 | –21.4 | 4.12 | –50.4 |
| $I_5$, $10^{-6}$·m$^{-1}$ | 31.6 | 13.1 | 22.6 | 4.05 |

## V. CONCLUSION

Simple expressions for the minimum emittance and corresponding initial horizontal beta and dispersion in the TME cell magnet with linear radius ramp are presented. Power series expansion with respect to the bending angle directly gives the uniform magnet parameters as a zero-order term while the field variation terms are superimposed over them. The logarithmic ratio of the peak central field to the edge one is the only parameter defining the emittance reduction factor below the uniform magnet.

The minimum emittance lattice cells designed confirm the theoretical predictions.

## ACKNOWLEGEMENT

The authors thank A.Bogomyagkov and K.Zolotarev for helpful discussions. This work was supported by Russian Foundation for Basic Research under Grant No. 15-02-04140 and by Russian Scientific Foundation Gran No. 14-50-00080.

## APPENDIX A: CLOSED-FORM MINIMUM EMITTANCE

We have obtained and compared the closed-form expressions for minimum emittance and for relevant initial horizontal beta and dispersion both with and without the focusing term $1/\rho^2(s)$. Even for rather strong peak field $B_c = B_{max}$ the difference is negligible, so here we ignore the focusing term. For brevity we introduce the parameter (see also Eq. (10))

$$y = \frac{k \cdot \theta}{2} = \ln\left(\frac{\rho_e}{\rho_c}\right) = \ln\left(\frac{B_c}{B_e}\right), \tag{A1}$$

where $k = \rho'(s)$ is the bending radius gradient Eq. (7) and $\theta$ is the total bending angle Eq. (9), and intermediary functions

$$A(y) = (4y - 23)e^{4y} + 8(y+5)e^{3y} - (y^2 + 12y + 10)e^{2y} - 8e^y + 1,$$

$$B(y) = 2e^{2y} - y^2 - 2y - 2.$$

Then the minimum emittance for the magnet with linear radius ramp is given by

$$\varepsilon_{x0v} = \frac{\theta^3}{12\sqrt{15}} \cdot \frac{3\sqrt{15} \cdot e^{-y}}{y^3(e^y - 1)} \cdot \left[\frac{A(y)B(y)}{e^{2y} - 1}\right]^{1/2}, \tag{A2}$$

with the following optimal beta and dispersion at the magnet midpoint

$$\beta_{x0v} = \frac{\rho_c \theta}{2\sqrt{15}} \cdot \frac{2\sqrt{15}}{y \cdot \sqrt{e^{2y} - 1}} \cdot \left(\frac{A(y)}{B(y)}\right)^{1/2}, \tag{A3}$$

$$\eta_{x0v} = \frac{\rho_c \theta^2}{24} \cdot \frac{12(e^{2y} - 4e^y + y + 3)}{y^2(e^{2y} - 1)}. \tag{A4}$$

In the uniform magnet limit as the radius gradient approaches 0, the above equations converge to the corresponding Eqs. (4, 5)

$$\eta_{x0v} \xrightarrow{k \to 0} \eta_{x0u} = \frac{\rho_c \theta^2}{24}, \quad \beta_{x0v} \xrightarrow{k \to 0} \beta_{x0u} = \frac{\rho_c \theta}{2\sqrt{15}}, \quad \varepsilon_{x0v} \xrightarrow{k \to 0} \varepsilon_{x0u} = \frac{\theta^3}{12\sqrt{15}}.$$

Note that the uniform TME bend parameters correspond to the peak field of the varied bend $B_c$, so the length of the uniform magnet is $L_u = \rho_c \cdot \theta$.

Fig.A1 plotted Eqs. (A2-A4) as a function of $y$ normalized to the relevant uniform magnet TME (i.e., ratios $r_\varepsilon = \varepsilon_{xov}/\varepsilon_{x0u}$ and the same for the optimal beta and the dispersion).

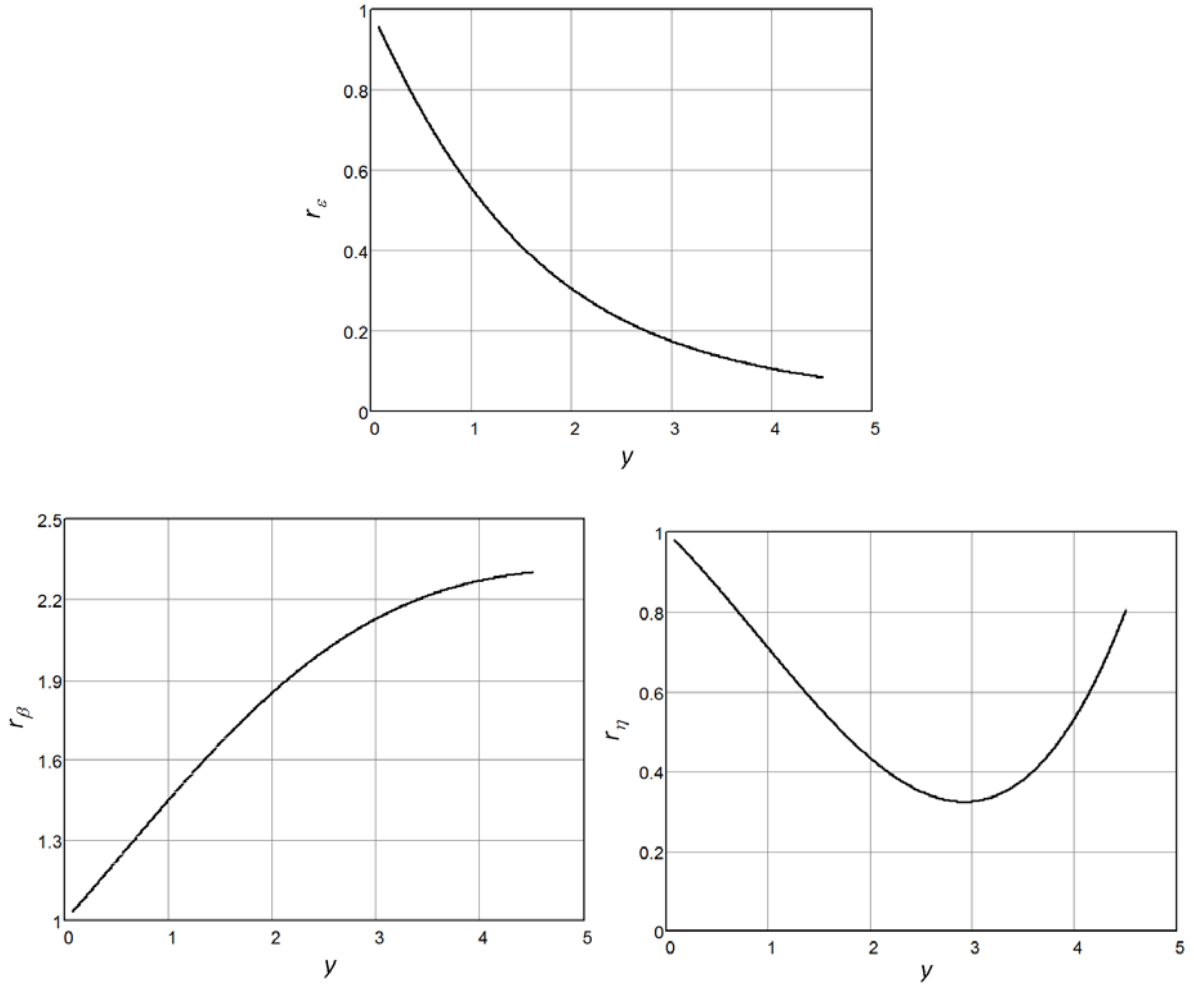

Fig.A1 Relative change in the minimum emittance (upper plot), initial beta (lower plot, left) and initial dispersion (lower plot, right) as a function of $y = \ln(B_c/B_e)$.